\RequirePackage{fix-cm}
\documentclass[sigconf,screen]{acmart}

\copyrightyear{2025}
\acmYear{2025}
\setcopyright{acmlicensed}\acmConference[FSE '25]{ACM International Conference on Foundations of Software Engineering }{June 23-June 25, 2025}{Trondheim, Norway}
\acmBooktitle{Companion Proceedings of the 33rd ACM Symposium on the Foundations of Software Engineering (FSE '25), June 23--27, 2025, Trondheim, Norway}
\acmDOI{10.1145/XXXX.XXXXXX}
\acmISBN{XXX-X-XXXX-XXXX-X/YY/MM}

\usepackage{graphicx}
\usepackage{textcomp}
\usepackage[utf8]{inputenc}
\usepackage{graphicx}
\usepackage{tabularx}
\usepackage{longtable}
\usepackage{hhline}
\usepackage{comment}
\usepackage{hyphenat}
\usepackage{subcaption}
\usepackage{float}
\usepackage{caption}
\usepackage{fancybox}
\usepackage{array,booktabs}
\usepackage{marvosym} 
\usepackage{textcomp} 
\usepackage{tablefootnote} 
\usepackage{balance}
\usepackage{multirow}
\usepackage{soul} 

\usepackage{enumitem} 

\usepackage{tcolorbox} 

\usepackage{listings} 
\usepackage{url}
\usepackage{hyperref}
\usepackage{adjustbox}
\usepackage{tcolorbox}
\usepackage{listings}
\usepackage{color}
\usepackage{colortbl}
\usepackage{pgfplots}
\usepackage{xspace}
\usepackage[noend]{algpseudocode}
\usepackage{algorithm}

\newcommand{\phead}[1]{\noindent{\bf #1}}

\newcommand{\rqboxc}[1]{\begin{tcolorbox}[left=4pt,right=4pt,top=4pt,bottom=4pt,colback=gray!5,colframe=gray!40!black,before skip=8pt,after skip=5pt]#1\end{tcolorbox}}

\makeatletter

\begin{document}
\title{Empowering AIOps: Leveraging Large Language Models for IT Operations Management}

\author{Arthur Vitui}
\authornote{Arthur Vitui is a Senior AI Specialist Solutions Architect at RedHat Canada.}
\affiliation{
  \institution{RedHat Inc.}
  \city{Montreal}
  \state{Quebec}
  \country{Canada}
}
\email{avitui@redhat.com}
\author{Tse-Hsun (Peter) Chen}
\affiliation{
  \institution{Software PErformance, Analysis and Reliability (SPEAR) Lab \\Concordia University}
  \city{Montreal}
  \state{Quebec}
  \country{Canada}
}
\email{ peterc@encs.concordia.ca}


\begin{abstract}
The integration of Artificial Intelligence (AI) into IT Operations Management (ITOM), commonly referred to as AIOps~\cite{GartnerAIOps}, offers substantial potential for automating workflows, enhancing efficiency, and supporting informed decision-making. However, implementing AI within IT operations is not without its challenges, including issues related to data quality, the complexity of IT environments, and skill gaps within teams. The advent of Large Language Models (LLMs) presents an opportunity to address some of these challenges, particularly through their advanced natural language understanding capabilities. These features enable organizations to process and analyze vast amounts of unstructured data, such as system logs, incident reports, and technical documentation. This ability aligns with the motivation behind our research, where we aim to integrate traditional predictive machine learning models with generative AI technologies like LLMs. By combining these approaches, we propose innovative methods to tackle persistent challenges in AIOps and enhance the capabilities of IT operations management.


\end{abstract}

\maketitle

\section{Introduction}


The incorporation of Artificial Intelligence (AI) into IT Operations Management (ITOM), commonly known as AIOps~\cite{GartnerAIOps}, offers considerable potential to automate processes, increase operational efficiency, and support better decision-making. AIOps is crucial in managing modern IT environments, which are increasingly complex, dynamic, and diverse~\cite{chen2017analytics, cheng2023ai, LWAKATARE2020106368, goswamichallenges}. Traditional monitoring and management approaches struggle to handle the sheer volume of data generated by logs, events, and metrics in real-time~\cite{cheng2023ai, Levin_2019, LWAKATARE2020106368, goswamichallenges, kumar2022challenges}. AIOps addresses these challenges by leveraging machine learning and AI to automate processes, detect anomalies, predict potential issues, and provide actionable insights. 

 However, integrating AI into IT operations presents numerous challenges that organizations must overcome to fully realize its potential benefits. Traditional AIOps is particularly difficult due to the complexity involved in consolidating diverse data sources, the need for significant feature engineering, and the dependence on narrowly focused machine learning models that require extensive domain expertise~\cite{LWAKATARE2020106368, goswamichallenges, kumar2022challenges}. Furthermore, many AI models, particularly those leveraging deep learning techniques, are often perceived as ``black boxes'' due to their lack of explainability~\cite{hedström2023quantusexplainableaitoolkit, kohlbrenner2020bestpracticeexplainingneural, LWAKATARE2020106368, goswamichallenges, kumar2022challenges}. In IT operations, understanding the rationale behind AI-driven insights is critical for fostering trust, maintaining compliance, and making well-informed decisions. Without interpretability, IT professionals may be reluctant to adopt AI recommendations, as they may not fully trust or comprehend the basis for these suggestions. 
 The above list is not exhaustive, as other factors can also impede the adoption of AI in IT operations. These include concerns about security and privacy, ethical and legal challenges, and issues related to cost and resource allocation. Addressing these considerations is essential to ensure that AI solutions are deployed responsibly and effectively~\cite{LWAKATARE2020106368, goswamichallenges, kumar2022challenges}.

 The emergence of Large Language Models (LLMs) presents a promising approach to addressing several challenges in IT Operations Management, particularly those involving data quality, the complexity of IT environments, and skill gaps. LLMs' advanced natural language understanding capabilities 
opens up new possibilities for data (including documentation) and problem analysis. Furthermore, by employing frameworks like ReAct~\cite{yao2023reactsynergizingreasoningacting}, IT operations professionals can gain insights into the LLM’s chain of thought~\cite{chu2023survey}, making it easier to understand how the AI arrived at a conclusion and generated an answer. These capabilities drive our research, where we explore the integration of tools including traditional predictive machine learning models with generative AI models such as LLMs. Our objective is to propose effective LLM-powered AI assistants for IT Operations Management, capable of addressing the aforementioned AIOps challenges. Additionally, we provide guidelines for evaluating these assistants, considering various business scenarios and requirements.

 In this paper, we share our experience deploying LLM-powered agents equipped with various tools to assist in tasks specific to managing a Kubernetes~\cite{kubernetes}-based environment, focusing on the Red Hat OpenShift~\cite{openshift} platform. These tools address diverse management challenges that IT operations professionals encounter daily. We leverage Retrieval Augmented Generation (RAG)~\cite{Edge2024FromLT} for summarizing and extracting procedures from platform documentation. Additionally, we developed platform analysis tools built on the Kubernetes API and other APIs, such as the Prometheus~\cite{prometheus} API, to extract both general-purpose and specific information related to platform and application management. These tools range from simple utilities like time conversion tools to more complex information extraction tools, such as retrieving KPI values from time-series data sources like Prometheus. Furthermore, we demonstrate the integration of predictive ML capabilities with LLMs by incorporating MLASP~\cite{MLASP}, a tool for capacity planning operations for a generic business application.

 We evaluate several models in a ten-fold execution loop over a set of 25 distinct tasks. These tasks are categorized into simple reasoning tasks, where the LLM responds to the input query using its training data or by utilizing at most one available tool, and advanced reasoning tasks, which require the LLM to identify and chain multiple tools (at least two) in the correct order to create a workflow for resolving the user query. In some cases, a specific tool may need to be used multiple times alongside others to achieve the correct solution. The tasks include general knowledge questions (e.g., introducing itself, answering general knowledge queries), time tracking questions commonly encountered in IT operations management (where the LLM must use the time tool to respond), platform and application information queries (e.g., retrieving version or deployment details), and data retrieval tasks (e.g., fetching specific KPIs within a time interval and presenting the results as a plot or a CSV table for reporting or further analysis). Additional details about these tasks are provided in Section~\ref{sec:llm-evaluation}. To address these tasks, the LLM agents employ the ReAct~\cite{yao2023reactsynergizingreasoningacting} principle, which integrates reasoning and action capabilities. Further information about this methodology is available in Section~\ref{sec:AIOPSFramework}.


 In summary, the paper makes the following contributions:
\begin{enumerate}
    \item We discuss typical challenges for adopting AIOps in industrial large scale environments and discuss the benefits of adopting LLM powered agents that use tools to solve typical ITOM tasks in Kubernetes environments. The benefits extend and may be used also on other types of large-scale deployments (virtualized or bare metal).
    \item We review the capabilities of different LLM models in solving ITOM tasks. We emphasize our results on the accuracy and cost differences (time to respond and token count) between these models.    
    \item We offer recommendations for selecting suitable LLMs and tools tailored to address specific business needs and challenges. Additionally, we outline directions for future research to address the current limitations identified through our experiments, aiming to enhance the integration and effectiveness of LLM-powered solutions in IT operations management.
\end{enumerate}

 We hope that our findings provide valuable insight to both researchers and practitioners in tackling the agentic assistant approach to AIOps methodologies. 

\phead{Paper organization.} 
Section \ref{sec:caseStudy} describes the studied system and the case study setup. Section \ref{sec:AIOPSFramework} describes our methodology. Section \ref{sec:results} presents the results. Section \ref{sec:discussions} discusses the lessons learned and our experience in the Kubernetes environment. Section \ref{sec:threats} discusses threats to the validity. Section \ref{sec:relatedWork} discusses related work. Section \ref{sec:conclusions} concludes the paper.

\section{Studied Systems}
\label{sec:caseStudy}

In this section, we present the studied system and the metrics we use to measure the performance of the LLM agents. 

\subsection{Studied Open Source System}
\label{sec:StudiedSystem-OpenSource}
We conduct our experiments on an open-source enterprise-grade Kubernetes platform, specifically Red Hat OpenShift~\cite{openshift}, a state-of-the-art Kubernetes environment and Platform as a Service (PaaS). OpenShift enhances Kubernetes with additional features that simplify and improve the development, deployment, and management of containerized applications. It provides developers with an integrated suite of tools, including built-in CI/CD pipelines, automated builds, and application catalogs, which streamline the entire application lifecycle. The platform also offers advanced security capabilities, such as integrated authentication, authorization, and policy management, ensuring compliance with enterprise-grade standards. Moreover, OpenShift supports hybrid and multi-cloud deployments, enabling organizations to run applications consistently across both on-premises and cloud environments. Its comprehensive ecosystem and developer-centric design make it an excellent platform for accelerating the delivery of modern, cloud-native applications.

Given OpenShift's extensive list of features, its management and the management of any hosted business application may require significant effort. For this reason, we evaluate the efficiency of IT operations management procedures using LLM-powered assistants with supporting tools.
We implement our agent-based solution to assist in such management. We host a custom workload on OpenShift that generically models various business applications using a black-box approach, capturing their typical characteristics: receiving a request, processing it (either internally or by calling other services), and returning a response. This custom workload is implemented using a generic application mocking framework called WireMock ~\cite{wiremock}, which allows us to simulate many requests. The complete details, features, and source code of the custom workload are available in the following GitHub repository~\cite{wiremock-metrics2}. From an operational standpoint, we want to manage the configuration and capacity planning of the mock business application to achieve a certain key performance indicator value (e.g., response time).


\subsection{Evaluation Metrics for Large Language Models}
\label{sec:metrics-llm}


 According to Sai et al.~\cite{sai2020surveyevaluationmetricsused}, the evaluation of LLMs should align with the specific tasks they are designed to perform (e.g., classification, question answering, summarization). Consequently, a wide variety of metrics may be applied depending on the task. They also emphasize the importance of human expert evaluation, particularly for open-ended generation tasks, where outputs should be rated based on fluency, coherence, relevance, and appropriateness. These human evaluations are crucial as quantitative metrics alone may fail to capture the subtleties of such tasks. Additionally, system performance metrics, such as response time and resource usage, must also be considered. Given the contextual nature of these evaluations, in the next sections, we detail the metrics applied in our research to assess the performance of LLM agents in the AIOps context:
 
\noindent\textbf{Compliance with Instructions}: This metric evaluates how effectively the model follows user instructions or prompts. For agents that utilize tools, this metric assesses the LLM's ability to \textit{determine which tools are required and the sequence in which they should be used}. It provides insight into the model's capacity to understand complex instructions and execute tasks accurately within the context. 

\noindent\textbf{Accuracy}: This metric calculates the ratio of correctly predicted instances to the total number of instances. While traditionally applied to classification tasks, in our context, this metric is used to evaluate how often the agent successfully completes a given task and provides the correct answer when asked repeatedly. The accuracy of the answer is determined through human expert evaluation, ensuring that the agent's outputs meet the expected standards. 

\noindent\textbf{Latency and Throughput}: This metric evaluates the time required for the model to generate responses and the number of responses it can produce within a given time frame. It is particularly significant for real-time applications where responsiveness is a critical factor. Ensuring low latency and high throughput is essential for scenarios that demand immediate interaction or decision-making, making this metric a key consideration in assessing the practical usability of LLMs in time-sensitive environments. 

\noindent\textbf{Cost}: This metric assesses the number of tokens generated by the LLM during the entire process of addressing a user request. It includes tokens used in reasoning through the chain of thought, executing the necessary steps to reach a conclusion, and formatting the final response. Monitoring token usage is crucial for understanding the model's efficiency and cost-effectiveness, particularly in scenarios where token limits or financial costs are tied to the usage of the LLM.

\section{The AIOPs Framework}
\label{sec:AIOPSFramework}
In this section, we describe our methodology for conducting AIOps-oriented experiments using an LLM-powered agentic approach. The core concept involves developing an integrated chatbot that IT operations professionals can engage with to streamline their daily tasks. Below, we delve into the integration of LLMs and the utilization of available tools within the target system to execute various operational activities effectively.

\subsection{Integration Overview}
\label{sec:ocp-integration}
Our evaluation system is the RedHat OpenShift~\cite{openshift} platform, a Kubernete-based PaaS frameworks~\cite{kubernetes}, combined with a custom business application modeled as a black-box using the WireMock~\cite{wiremock} testing framework. Applications in OpenShift are organized within namespaces to group or isolate them based on security requirements. These applications can operate as standalone components or be managed by operators, which automate tasks like creation, configuration, and management of Kubernetes-native (Knative) application instances. The ecosystem includes operator-managed applications, standalone applications, and platform baseline tools supporting business Knative applications, security, and lifecycle management.

Our chatbot, a Knative application, interfaces with IT operations professionals, leveraging an LLM and specialized tools to handle requests. Key components of this chatbot include: (1) A user interface for input, bypassed in our testing as we used predefined queries. (2) An LLM client module to interact with an LLM inference server. Optional memory components can enable conversational capabilities but may interfere with reasoning (see Section~\ref{sec:discussions}). We utilized LangChain~\cite{langchain} and LangGraph~\cite{langgraph} for this integration. (3) A set of tools integrated with the LLM client via LangChain, LangGraph, and Pydantic~\cite{pydantic}. Pydantic defines tool interfaces for the LLM, specifying expected inputs, outputs, and formatting instructions to ensure seamless interaction and result interpretation.

For our experiments, we created a list of custom-made tools, $T<n>$, with $n\in [1,9]$, built using the Python programming language and the supporting libraries. The purpose of these tools is to help in various ITOM tasks such as capacity planning, procedure summary extraction, platform deployment and configuration information, platform and application KPI extraction for a defined datetime range with CSV and graphical outputs. In Section ~\ref{sec:llm-framework}, we describe how the LLMs utilize these tools to respond to user queries. In Section~\ref{sec:llm-evaluation}, we describe the user queries we created to perform our tests alongside the tool or the list of tools the LLMs must use to correctly respond to queries. Next, we name the tools and briefly describe them.



\textbf{T1, MLASP~\cite{MLASP}:} is a machine learning based capacity planning tool that generates a set of parameter configurations (e.g., size of a thread pool) to support a desired KPI value for the WireMock application. 

\textbf{T2, RAG:} a tool that gives the LLM the ability to search a specialized vector database that contains encoded documentation about the Red Hat OpenShift AI operator, including procedure description and how-to information. The LLM can inspect this database to obtain information based on the received query and then summarize a response to the user. 

\textbf{T3, Time info:} a tool that calculates the timestamp, the iso-formatted string, and the timezone string of the requested time information. Returns a Python object containing the timestamp value, the ISO formatted string of the date-time value, and the timezone string. 

\textbf{T4, List operators:} is used to find information about operators installed within a namespace. The response may contain information such as the name of the operator, its version, and deployment status. 

\textbf{T5, Pod summary:} is used to summarize information about the pods that exist in a namespace. A pod is a Kubernetes abstraction that represents a group of one or more application containers and some shared resources for those containers (e.g., shared storage, networking cluster, etc.). The tool returns an object containing the name of the namespace and pod state (e.g., running, stopped) and pod counter information. For the running pods, it also returns its name and, if available, any service information such as service name, service ports, and route. 

\textbf{T6, Service summary:} is used to summarize service information in an OpenShift namespace. It returns an object containing the name of the namespace and a list of the available services and their properties, such as name, port numbers, and route information.

\textbf{T7, Prometheus metric names:} list available metric names in a Prometheus~\cite{prometheus} instance using an input filter. Prometheus is an open-source technology designed to provide monitoring and alerting functionality for cloud-native environments, including Kubernetes. The inputs for this tools are a filter name and value for the filter (e.g. input filter name is 'namespace' and filter value is 'demo'). It returns a list containing the available metric names. 

\textbf{T8, Prometheus metric data range:} is used to list the application metric values and associated timestamps between a start and an end timestamp interval for a given metric name stored within a Prometheus instance. It returns a pydantic object containing the list of the desired application metric values and associated timestamp information. 

\textbf{T9, Plot prometheus metric range data as file:} is used to create a file with the plot of the instantaneous rate (irate) of an application metric values and associated timestamps between a start and an end timestamp interval for a given metric name stored within a Prometheus instance. It returns a string containing the name of the file containing the plot.

The tools listed above enable us to test a limited range of IT Operations Management (ITOM) operations and scenarios. However, the use cases they support are sufficient to evaluate and report on the performance of various LLMs. Naturally, this toolset can be expanded to incorporate additional tools, thereby broadening the ITOM management capabilities of the LLM-powered AI assistant and supporting a wider array of operational tasks.

\subsection{ReAct Large Language Model Agents}
\label{sec:llm-framework}
The AI assistant for ITOM tasks is powered by the ReAct~\cite{yao2023reactsynergizingreasoningacting} (Reason and Act) framework. ReAct enhances Large Language Model (LLM)-based agents by integrating reasoning and action tasks into a unified approach. These agents combine the natural language understanding and generation capabilities of LLMs with the ability to interact with external tools and environments, enabling them to provide sophisticated, context-aware, and actionable responses. ReAct-based agents excel in analyzing problems, performing necessary actions (e.g., querying a database or using an API), and refining their understanding based on action outcomes. This integrated reasoning and acting approach employs chain-of-thought prompting~\cite{wei2023chainofthoughtpromptingelicitsreasoning}, allowing agents to generate intermediate reasoning steps. This enhances reasoning transparency and significantly improves their ability to manage complex, multi-step tasks. Most importantly, ReAct agents utilize external tools, APIs, and databases to access real-time information and perform computations beyond their pre-trained knowledge. This capability extends the agent's functionality, enabling accurate, timely, and actionable responses that are critical for IT operations management.

 To build our agent, we use the standard ReAct instructions provided in the LangGraph~\cite{langgraph} library. The LangChain~\cite{langchain} and LangGraph frameworks offer a high-level abstraction for interacting with various large language models (LLMs) across different providers and model variants. This abstraction ensures consistency during testing by accommodating the unique prompt formats and data-passing requirements of different LLMs. Leveraging these frameworks eliminates the need for custom integration code for each model, simplifying the experimentation process. To maintain consistency and avoid introducing potential biases that could influence the behavior of the evaluated LLMs, we rely on the default prompts provided for each model.

\begin{verbatim}
Answer the following questions as best you can. 
You have access to the following tools:

{tools}

Use the following format:

Question: the input question you must answer
Thought: you should always think about what to do
Action: the action to take, should be one of 
[{tool_names}]
Action Input: the input to the action
Observation: the result of the action
... (this Thought/Action/Action Input/Observation 
can repeat N times)
Thought: I now know the final answer
Final Answer: the final answer to the original input 
question

Begin!

Question: {input}
Thought:{agent_scratchpad}
\end{verbatim}


 This agent-driven strategy improves operational efficiency by minimizing the cognitive load on IT teams, enabling quicker issue resolution, and supporting seamless continuous deployment and application scaling.  Ultimately, this integration facilitates a more intelligent, agile, and scalable IT infrastructure that drives better business outcomes.

\subsection{Evaluating LLM Powered Agents in AIOps Context}
\label{sec:llm-evaluation}

\begin{figure}
	\centering
	\includegraphics[width=0.9\linewidth]{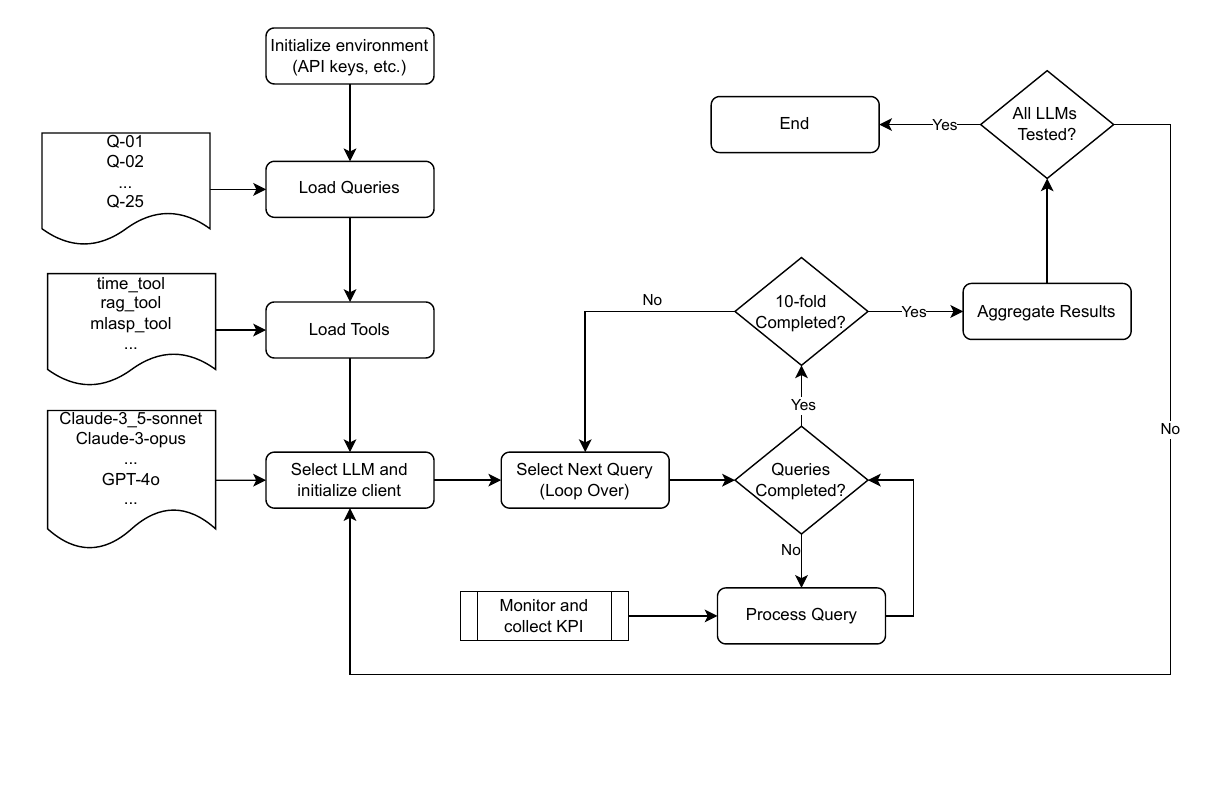}
	\caption{An example workflow for LLM performance evaluation in an AIOps context.}
	\label{fig:aiops-evaluation-flow}
\end{figure}

The overall evaluation process is illustrated in Figure~\ref{fig:aiops-evaluation-flow}. We evaluate the capabilities of different LLMs to act as AI assistants for ITOM tasks by asking them to respond to a list of queries detailed in Table~\ref{tab:llm-questions}. The queries are a mix of general-purpose queries (e.g., Q-01, Q-02, Q-08), specific platform-related queries (e.g., Q-05, Q-10, Q-13), and target application management questions (e.g., Q-21, Q-23, Q-24). In order to respond to these queries, the LLMs must use their training data, or the available list of tools. Additionally, we categorize these queries into two distinct types: \textbf{Simple Reasoning (SR)}: where the LLM must respond by relying solely on its training data or, at most, utilizing one tool; and \textbf{Advanced Reasoning (AR)}: where the LLM must identify multiple tools to use and construct a workflow in which the tools are employed in the correct sequence. In advanced reasoning scenarios, the LLM is also responsible for formatting the data as required before using a tool, and the same tool may be invoked multiple times with varying inputs.

 We test these questions on multiple large language models and various variants of the same model, assessing their responses for accuracy and collecting key performance metrics such as response time and the average number of tokens consumed to produce an answer. To evaluate each model's robustness and consistency, we ask each model the same question 10 times. We evaluate our queries on the following list of commonly-used LLMs: from the Anthropic family: Claude 3.5 Sonnet, Claude 3 Haiku, and Claude 3 Opus; from the Mistral family: Mistral Largest, Mixtral 8x22B, and Mistral Small 7B; from the OpenAI family: GPT 3.5 Turbo, GPT 4-o, GPT 4-o Mini, and GPT 4 Turbo.

\begin{table}
    \centering
    \caption {Large Language Model Agents for AIOps Evaluation Queries. Cat. shows the category of the question, where SR means Simple Reasoning (use at most one tool) and AR means Advanced Reasoning (use multiple tools). }
    \scalebox{0.88}{
    \begin{tabular}{|p{0.036\textwidth}|p{0.024\textwidth}|p{0.035\textwidth}|p{0.34\textwidth}|}
        \hline
        \textbf{Q\#.} & \textbf{Cat.} & \textbf{Tools} & \textbf{Query Text}\\
        \hline
        Q-01 & SR & - & Hi, who are you? \\
        \hline
        Q-02 & SR & - & What tools do you have access to? \\
        \hline
        Q-03 & SR & - & Give me the list of tools you have access to. \\
        \hline
        Q-04 & SR & - & Give me the list and a short description of the tools you have access to.\\ 
        \hline
        Q-05 & SR & T4 & What operators are in namespace demo? \\
        \hline
        Q-06 & SR & T4 & What operators are in namespace demo? Please provide only the name and the version for each operator.\\
        \hline
        Q-07 & SR & T2 & How can I create a Data Science Project? \\
        \hline
        Q-08 & SR & - & Can you describe Paris in 100 words or less? \\
        \hline
        Q-09 & SR & - & Is there a river? \\
        \hline
        Q-10 & SR & T5 & Tell me about the pods in namespace demo. \\
        \hline
        Q-11 & SR & T5 & Give me a summary of the running pods in namespace demo. Please include service and route information in the response. \\ 
        \hline
        Q-12 & SR & T5 & Give me the complete summary of the pods in namespace demo. \\
        \hline
        Q-13 & SR & T5 & Give me a summary of the running pods in namespace demo. Give me only the names and the route if they have one.\\
        \hline
        Q-14 & SR & T3 & What day is today? \\
        \hline
        Q-15 & SR & T3 & What is the current date time? \\
        \hline
        Q-16 & SR & T3 & What is the current timestamp? \\
        \hline
        Q-17 & SR & T3 & What is the timestamp and date time for 3 hours ago? \\
        \hline
        Q-18 & SR & T3 & What is the timestamp and date time for 3 hours from now? \\
        \hline
        Q-19 & SR & T3 & What is the timestamp and date time for 3 hours ago? \\
        \hline
        Q-20 & SR & T6 & Is there a prometheus service running in namespace demo? If so, give me its name and port values. \\
        \hline
        Q-21 & AR & T6, T7 & Find out the service name and port number of the Prometheus service running in namespace demo. Then use that information to retrieve the list of metrics filtered by namespace demo.\\
        \hline
        Q-22 & AR & T6, T7 & Find out the Prometheus service name and port number running in namespace demo. Give me all the metrics stored by it that have a name that starts with load\_generator.\\
        \hline
        Q-23 & SR & T1 & What configuration of WireMock supports a throughput KPI of 307 within a 2.9 percent precision? Search for 100 epochs to find the result.\\
        \hline
        Q-24 & AR & T3, T6, T9 & Find out the Prometheus service name and port number running in namespace demo. Use it to to plot all the prometheus metric data for the metric load\_generator\_total\_msg starting 40 days ago until now. Return only the file name and nothing else.\\
        \hline
        Q-25 & AR & T3, T6, T8 & Find out the Prometheus service name and port number running in namespace demo. Use that to get all the prometheus metric data for the metric load\_generator\_total\_msg starting 40 days ago until now. Print out only the metric values and their associated timestamp as a CSV table.\\
        \hline
    \end{tabular}}
    \label{tab:llm-questions}
\end{table}

\section{Case Study Results}
\label{sec:results}
In this section, we discuss the results of the different LLM powered agents' performance in an AIOps setting. Due to space constraints, we only discuss the overall results, while the full results for each individual question are available online~\cite{llm-eval-aiops-code}. 
The experimentation approach applicable to all our research questions is the following: we perform the procedure described in Section ~\ref{sec:llm-evaluation} and depicted in Figure~\ref{fig:aiops-evaluation-flow} for each LLM mentioned and each query formulated in that section.

\subsection{RQ1: How accurately do different LLMs perform on a set of IT Operations tasks?}
\label{sec:aiops_rq1}
\noindent {\bf Motivation:} 
Large Language Models are different from one another from multiple perspectives, such as training data, architecture, and number of parameters, which can affect the abilities on performing IT Operations tasks. 
Given these differences we want to evaluate how accurately different models perform on ITOM tasks.

\noindent\textbf{Results:} Table~\ref{tab:aiops-rq1-results} summarizes the accuracy of the models we tested, following the outlined process in our approach. We notice that some models perform better on simple reasoning tasks, some on the advanced reasoning ones, while some perform similarly well on both type of tasks. We find that the Anthropic models family is the best in class for simple reasoning and OpenAI's models are better on advanced reasoning. Looking at the lower level details for advanced reasoning, the Claude 3.5 Sonnet and Claude 3 Opus have high accuracy rates for advanced reasoning, however the answers they provided there lacked the details and completeness provided by the GPT-4 models. 

Surprisingly, GPT 3.5 was not able to correctly respond to any of the advanced reasoning tasks as it was not able to create the correct workflow and tool chaining required by the query. We also note that within the Mistral family of models, the mixture of experts model Mistral 8x22B hallucinated most responses. Given that this model is closed source, we cannot determine the specific reasons behind its suboptimal behaviour. We also notice that despite their reduced size, smaller models can still demonstrate notable efficiency in reasoning and selecting appropriate tools to complete tasks, as was the case of Mistral Small 7B. Due to space constraints we only provide the summary, however, the full results, for each question, are available in the GitHub repository~\cite{llm-eval-aiops-code}. Additionally, further details and explanations regarding various aspects of model responses and their accuracy are discussed in Section~\ref{sec:discussions}. 

\begin{table}
    \centering
    \caption {RQ1 - Large Language Model Agents solving task accuracy in AIOps Context.}
    \begin{tabular}{|c|c|c|}
        \hline
        \textbf{Model} & \textbf{SR Task} & \textbf{AR Task}\\
        & (0..1 tools) & (2+ tools) \\
        \hline
        Claude 3.5 Sonnet & 95.23\% & 95\% \\
        \hline
        Claude 3 Haiku & 89.52\% & 30\% \\
        \hline
        Claude 3 Opus & 90\% & 95\% \\
        \hline
        Mistral Largest & 94.76\% & 45\% \\
        \hline
        Mixtral 8x22B & 2.38\% & 0\% \\
        \hline
        Mistral Small 7B & 76.19\% & 0\% \\
        \hline
        GPT 3.5 Turbo & 85.71\% & 0\% \\
        \hline
        GPT 4-o& 87.14\% & 100\% \\
        \hline
        GPT 4-o Mini & 85.71\% & 77.5\% \\
        \hline
        GPT 4 Turbo & 85.71\% & 90\% \\
        \hline
    \end{tabular}
    \label{tab:aiops-rq1-results}
\end{table}

\rqboxc{Except for Mixtral 8x22B, all tested LLMs adhered to the ReAct principles and effectively utilized the available tools when the instructions were unambiguous. Among the models, the GPT family demonstrated the best overall performance for advanced reasoning, with GPT 4-o leading in accuracy and reliability, while the Claude model family performed best on simple reasoning. Conversely, Mixtral 8x22B exhibited the worst performance, frequently hallucinating responses, and did not incorporate the tools provided.}

\subsection{RQ2: How fast do different LLMs perform on a set of IT Operations tasks?}
\label{sec:aiops_rq2}
\noindent\textbf{Motivation:} 
Building on the previously discussed differences among LLMs, we evaluate how quickly various models respond to user requests. Specifically, we record the 50th percentile (median), 90th percentile, and maximum response times (in seconds) for models differing in size, architecture, context window, and training data. These measurements are crucial for tasks requiring quick responses, such as real-time applications in chatbots, virtual assistants, customer support systems, and preventive maintenance workflows. Delayed responses can lead to user frustration, decreased engagement, and perceptions of system unreliability. 

Evaluating response times at the 50th percentile (median) and 90th percentile offers a comprehensive view of performance. The 50th percentile reflects the typical response time, experienced by half of the users, representing general system efficiency. The 90th percentile highlights the response time below which 90\% of requests are completed, revealing tail-end delays that can significantly impact user satisfaction. By analyzing these metrics, developers can identify latency bottlenecks and optimize system performance to deliver consistent, timely responses for a majority of users, ensuring both reliability and effectiveness.


\noindent\textbf{Results:}  In Table~\ref{tab:aiops-rq2-results}, we summarize the performance metrics observed as part of this research question. Due to space restrictions, we report the average time taken to respond in the class of tasks (simple reasoning and advanced reasoning); however, full details per query are available in the GitHub repository~\cite{llm-eval-aiops-code}.

 Response times should always be assessed in conjunction with the accuracy of the responses to provide meaningful insights. Evaluating these metrics together allows for a more comprehensive understanding of a model’s performance, particularly in scenarios where speed and correctness are both critical. Without this context, fast response times may overshadow inaccuracies, or highly accurate responses may not meet the time requirements for real-time applications. With this perspective, we find that for \textbf{SR queries}, OpenAI models generally respond the fastest, closely followed by Anthropic models. Notably, Claude 3 Haiku emerges as the fastest responder for SR queries based on P-50 response times. The Mistral family models, except for Mixtral 8X22B, also deliver competitive response times for SR tasks. For \textbf{AR queries}, the response times among models capable of solving these queries are relatively similar. 

 However, smaller models within a family generally respond about 50\% faster than their larger counterparts. When considering both response time and accuracy, OpenAI models, particularly GPT-4o, stand out as the fastest models offering the best responses. On the other hand, the Mistral models, particularly Mixtral 8x22B, perform the worst. While Mixtral 8x22B returned hallucinated responses quickly, it was unable to resolve any of the AR queries. However, for SR queries, the Mistral Small 7B model remains a viable option due to its overall accuracy as observed in RQ1. Balancing its relatively fast response times with reasonable accuracy could make it acceptable for specific ITOM operations scenarios where precision is not the primary concern.

We notice that larger models tend to take more time to respond to complex queries due to their larger parameter sizes, which enable them to handle and process more intricate tasks successfully. We further observe that response times are also influenced by other factors, such as the performance of the tools invoked during the process, the number of tools required to generate a response, the internal reasoning time of the LLM, and the payload transfer times between the LLM and the tools. These elements collectively contribute to the overall latency, and larger models often trade off response speed for their ability to manage greater task complexity and deliver accurate results.

\begin{table}
    \centering
    \caption {RQ2 - Large Language Model Agents solving task average response times (in seconds) in AIOps Context.}
    \begin{tabular}{|c|c|c|c|}
        \hline
        \textbf{Model} & \textbf{Metric} & \textbf{SR Task} & \textbf{AR Task}\\
        & (Average) & (0..1 tools) & (2+ tools) \\
        \hline
        Claude 3.5 & P-50 & 6.41 & 19.14 \\
        Sonnet & P-90 & 7.12 & 20.54 \\
        \hline
        Claude 3 & P-50 & 3.14 & 9.07 \\
        Haiku & P-90 & 4.38 & 16.08 \\
        \hline
        Claude 3 & P-50 & 18.18 & 48.38 \\
        Opus & P-90 & 20.97 & 55.12 \\
        \hline
        Mistral & P-50 & 8.36 & 71.95 \\
        Largest & P-90 & 12.31 & 88.04 \\
        \hline
        Mixtral & P-50 & 4.73 & 6.88 \\
        8x22B & P-90 & 5.72 & 7.17 \\
        \hline
        Mistral & P-50 & 4.72 & 9.12 \\
        Small 7B & P-90 & 5.17 & 9.42 \\
        \hline
        GPT 3.5 & P-50 & 3.45 & 6.99 \\
        Turbo & P-90 & 4.25 & 7.35 \\
        \hline
        GPT 4-o & P-50 & 4.47 & 46.12 \\
        & P-90 & 5.69 & 54.42 \\
        \hline
        GPT 4-o & P-50 & 4.13 & 22.08 \\
        Mini & P-90 & 6.06 & 28.61 \\
        \hline
        GPT 4 & P-50 & 9.17 & 48.62 \\
        Turbo & P-90 & 10.53 & 54.99 \\
        \hline
    \end{tabular}
    \label{tab:aiops-rq2-results}
\end{table}

\rqboxc{Response times must be evaluated in context and alongside the accuracy of the responses. In this context, and considering all response times, OpenAI models generally respond the fastest on both SR and AR queries. However, based on the P-50 metric values, for SR queries, Claude 3 Haiku emerges as the fastest model, where for AR queries, the GPT-4o is the fastest.}

\subsection{RQ3: How verbose do different LLMs perform on a set of IT Operations tasks?}
\label{sec:aiops_rq3}
\noindent\textbf{Motivation:} 
Building on the differences between LLMs mentioned earlier, we assess how verbose each model is when responding to user requests by calculating the average number of tokens used per response. This average is determined by repeating the same request 10 times and recording the token count for each response. 
Measuring verbosity is important because it impacts both response time—longer responses take more time to stream and require more effort for a human to process—and operational costs, especially for third-party-hosted models where usage is billed per token. Furthermore, a verbose model with a limited context window may face constraints in addressing complex queries or executing intricate workflows effectively.


\noindent\textbf{Results:} In Table~\ref{tab:aiops-rq3-results}, we present the verbosity results for various LLMs as they perform AIOps-related tasks. 
Similar to the observations in prior research questions, a model's verbosity is closely linked to its task-solving capability. Generally, larger models tend to provide more verbose answers. Additionally, verbosity varies between models from different providers, even when they belong to similar classes (e.g., Anthropic's Claude 3.5 Sonnet, Mistral's largest model, and OpenAI's GPT 4-turbo). This highlights provider-specific differences in how models approach and structure their responses.

In general, Anthropic's Claude family models exhibit the highest verbosity, even though their responses are not as complete or accurate as those of OpenAI's GPT models. For example, Claude models consistently truncated their responses to query Q-25, whereas GPT 4-turbo provided a complete response. The verbosity of the Mistral family models is often comparable to that of the Claude family, but only when correct responses were recorded. This highlights differences in verbosity that do not always correlate with response quality or completeness.

\begin{table}
    \centering
    \caption {RQ3 - Average token count (verbosity) of Large Language Model Agents solving tasks in AIOps Context.}
    \begin{tabular}{|c|c|c|}
        \hline
        \textbf{Model} & \textbf{SR Task} & \textbf{AR Task}\\
        & (0..1 tools) & (2+ tools) \\
        \hline
        Claude 3.5 Sonnet & 5420.3 & 42768.1 \\
        \hline
        Claude 3 Haiku & 5564.6 & 46535.7 \\
        \hline
        Claude 3 Opus & 5892.6 & 44648.5 \\
        \hline
        Mistral Largest & 6066.8 & 39877.6 \\
        \hline
        Mixtral 8X22B & 2803.4 & 3007.3 \\
        \hline
        Mistral Small 7B& 4162.7 & 4619.5 \\
        \hline
        GPT 3.5 Turbo & 3242.3 & 2722.6 \\
        \hline
        GPT 4-o& 3031.6 & 25965 \\
        \hline
        GPT 4-o Mini & 3065 & 20900.7 \\
        \hline
        GPT 4 Turbo & 3081.3 & 24308.5 \\
        \hline
    \end{tabular}
    \label{tab:aiops-rq3-results}
\end{table}

\rqboxc{Overall, the OpenAI model family demonstrates the most efficiency in token usage for both SR and AR queries, while the Anthropic models tend to be the most verbose. This distinction is critical as it directly impacts the operational cost of using the model for ITOM tasks. More verbose models can significantly increase overall costs, making them less appealing from a project cost management perspective.}  

\section{Discussions}
\label{sec:discussions}
In this section, we present the key observations gained from our evaluation of the tools we developed for the LLM agents and their integration into the Red Hat OpenShift~\cite{openshift} platform.


\subsection{Understanding AI Assistants Behavior With and Without A Memory Component}
\label{sec:discussion-memory-aspects}
Chat agents can be implemented with or without a memory component. Having a memory component retains aspects of earlier queries. While using a memory component may enhance the conversational experience by enabling follow-up queries and maintaining context, it introduced unexpected behavior in our experiments. For instance, it worked well for queries Q-08 and Q-09, where continuity was beneficial. However, for queries Q-15 through Q-25, the memory component caused incorrect responses, as the LLM failed to calculate updated timestamp information. To ensure accurate responses for real IT operations queries, we disabled the memory component. This adjustment allowed the agents to provide correct calculations but resulted in 100\% incorrect responses for Q-09, as the LLM could no longer infer context from Q-08 (e.g., the expected response for Q-09 would reference the river Seine).

Removing the memory component also addressed issues with models that have shorter context windows, as these windows are directly affected by the length of the chat history. Additionally, in many IT operations tasks, the responses generated (e.g., Q-24 returns a file name) are not reusable in subsequent queries, making the memory feature less practical. However, we recognize that, in certain scenarios, chat history can be valuable. For example, after solving Q-23, a user might want the agent to apply the response on the system using an available tool. Balancing these considerations, we opted to remove the memory component to prioritize correctness in task-oriented queries.

\subsection{Understanding Model Evaluations In Context}
\label{sec:discussion-context}
In the detailed responses for RQ1 available in the GitHub repository~\cite{llm-eval-aiops-code}, we noted that queries Q-04 and Q-21 were not successfully resolved by the GPT 3.5 turbo model due to an internal processing error reported by the integration library. The correlation between token count and query results is particularly evident in these two cases. In the same repository for details of RQ3, the token count is recorded as zero for these queries, even though the model took some time before returning an error, as reflected in the detailed response time metrics of RQ2~\cite{llm-eval-aiops-code}. This highlights a disconnect between the processing time and token generation in the case of internal model errors.

\subsection{Understanding Model Strengths and Weaknesses for Advanced Reasoning Tasks}
\label{sec:discussion-ar}

Some of the queries used in our tests to evaluate the agents' reasoning capabilities required the detection and correct sequential use of multiple tools, including repeated use of the same tool within a single workflow. For example, Table~\ref{tab:aiops-discussion-q24-correct} outlines the correct reasoning process and the steps taken by the LLM to solve such a task. The response was successfully generated by OpenAI's GPT 4-turbo, demonstrating its ability to handle complex workflows and execute tasks in the appropriate order.

\begin{table}
    \centering
    \caption {Query Q-24 correctly solved by GPT 4-turbo agent.}
    \scalebox{0.92}{
    \begin{tabular}{|c|p{0.42\textwidth}|}
        \hline
        \textbf{Step\#} & \textbf{Agent Activity}\\
        \hline
        1 & Input: Find out the Prometheus service name and port number running in namespace demo. Use it to to plot all the prometheus metric data for the metric load\_generator\_total\_msg starting 48 hours ago until now. Return only the content string of the tool and nothing else. \\
        \hline
        2 & Entering new AgentExecutor chain... To solve the question, I need to first identify the Prometheus service name and port number in the "demo" namespace. Then, I will use this information to plot the Prometheus metric data for "load\_generator\_total\_msg" from 48 hours ago until now. \\
        \hline
        3 & Action: Summarize\_Services\_Information\_In\_OpenShift\_Namespace \\
          & Action Input: \{"namespace": "demo"\} namespace = 'demo' svc\_summary = [ServiceInfo (name = 'grafana-demo-service', ports = [PortInfo (port=3000, name = 'grafana', protocol = 'TCP')], route = 'http://grafana-demo.apps.cluster-dnjmk.dnjmk.sandbox1590.opentlc.com/'), ServiceInfo (name = 'influxdb', ports = [PortInfo (port = 8086, name = 'No name available', protocol = 'TCP')], route = 'unavailable'), ServiceInfo (name = 'load-generator', ports = [PortInfo ( port = 9090, name = 'metrics-app', protocol = 'TCP'), PortInfo (port = 9100, name = 'metrics-node', protocol = 'TCP')], route = 'unavailable'), ServiceInfo (name = 'prometheus-operated', ports = [PortInfo (port = 9090, name = 'web', protocol = 'TCP'), PortInfo (port = 10901, name = 'grpc', protocol = 'TCP')], route = 'unavailable')]\\
        \hline
        4 & From the observation, the Prometheus service in the "demo" namespace is named "prometheus-operated" and it uses port 9090 for web access. Next, I need to get the current time and the time 48 hours ago to define the time range for plotting the metric data. \\
        \hline
        5 & Action: Get\_timestamp\_and\_time\_ISO \\
          & Action Input: \{"time\_value": "now", "time\_metric": "seconds", "ago\_flag": 0\} timestamp = 1730500568.411993 date\_time\_iso\_format\_string = '2024-11-01T18:36:08.411993-04:00' timezone = 'America/New\_York' \\
        \hline
        6 & Now I have the current timestamp. Next, I need to calculate the timestamp for 48 hours ago. \\
        \hline
        7 & Action: Get\_timestamp\_and\_time\_ISO \\
          & Action Input: \{"time\_value": 48, "time\_metric": "hours", "ago\_flag": 1\} timestamp = 1730327770.333979 date\_time\_iso\_format\_string = '2024-10-30T18:36:10.333979-04:00' timezone = 'America/New\_York' \\
        \hline
        8 & Now I have the timestamp for 48 hours ago. With both timestamps, I can proceed to plot the metric data for "load\_generator\_total\_msg" from 48 hours ago until now using the Prometheus service details. \\
        \hline
        9 & Action: File\_create\_plot\_irate \\
          & Action Input: \{"prom\_service": "prometheus-operated", "prom\_namespace": "demo", "prom\_port": 9090, "metric\_name": "load\_generator\_total\_msg", "metric\_range\_start": 1730327770.333979, "metric\_range\_end": 1730500568.411993\} file\_name='FILE-plot-load\_generator\_total\_msg-1730327770-1730500568.png' \\
        \hline
        10 & I now know the final answer \\
           & Final Answer: FILE-plot-load\_generator\_total\_msg-1730327770-1730500568.png \\
        \hline
    \end{tabular}}
    \label{tab:aiops-discussion-q24-correct}
\end{table}

The process begins with the agent receiving the input query in step 1. In step 2, the agent generates a brief summary of the reasoning approach needed to solve the task. In step 3, the agent identifies the initial tool required, which is the \textit{Summarize\_Services\_Information\_In\_OpenShift\_Namespace} tool. This tool is used to retrieve the Prometheus service name and port number for the Prometheus operator instance running in the specified namespace—\textit{demo}, in this case. The tool's output is a list of Pydantic objects, which the LLM parses and provides as an observation in step four. During step 4, the LLM also outlines the next steps to be taken. In step 5 through 8, the agent uses the \textit{Get\_timestamp\_and\_time\_ISO} tool twice to calculate the timestamps for the required time range. These timestamps are then utilized in step nine with the \textit{File\_create\_plot\_irate} tool to generate the plot. The execution of the tool in step 9 returns the plot as a file. Finally, in step 10, the agent executor reviews the results, concludes that the response is complete, and returns the file name as the final answer.

As illustrated in the detailed results on GitHub~\cite{llm-eval-aiops-code}, seven of the 10 models tested were able to respond to Q-24 with varying levels of precision, with six models achieving an accuracy rate greater than 80\%. Upon examining the chain call logs, the following reasons for model failures were identified: \textbf{Hallucinations:} The model hallucinates the date range instead of utilizing the tool to compute the correct time interval. \textbf{Query Deflection:} The model fabricates a response by suggesting what the user should do to obtain the answer, rather than directly addressing the query. This could also be considered a form of hallucination. \textbf{Flawed Reasoning:} The model fails to create the appropriate workflow and sequence of tools necessary to solve the task, ultimately responding that it could not find any relevant information.
In some instances, models from the Anthropic family fail to determine the correct order for utilizing tools (notably, the timestamp calculation tool). However, they often manage to correct these errors during execution. Although this ability to recover from erroneous reasoning is a positive trait, it comes at the cost of increased response time and higher token consumption. This inefficiency could make the approach economically impractical for operations teams.

Surprisingly, although Q-25 is nearly identical to Q-24 in terms of reasoning requirements, only GPT 4-turbo and GPT 4-o models managed to handle it effectively. Other models faced challenges in providing responses, which included: \textbf{Response Truncation:} This occurred in the case of Anthropic models. The impact of truncation depends on how the resulting information is used. If the output is solely for display or review, truncation may not be problematic. However, if the result is intended for further processing, such as input into a reporting tool for additional calculations, truncation should be considered a processing error. \textbf{Improper Tool Usage:} Models like Anthropic Claude 3 Haiku, OpenAI's GPT 3.5 turbo, and GPT 4-o mini struggled with correctly utilizing tools, particularly the timestamp calculation tool. \textbf{Response Hallucination:} This was observed in MistralAI's Mixtral 8x22B and Mistral Small models, where they fabricated answers instead of generating them through tool use. \textbf{Timeouts:} The Mistral Largest model failed to respond within the allocated time, leading to a timeout.

MistralAI offers an LLM variant in the Mixtral8 family that operates as a mixture of experts, with different weight variants. However, our experiment revealed that this model was the poorest performer. It frequently hallucinated responses to IT operations tasks and consistently failed to utilize the tools provided for calculations. Due to its closed-source nature, identifying the underlying cause of this behavior was not possible. We suggest future research to refine the agentic approach and re-assess the capabilities of this model, particularly with improved tuning and configuration.

In summary, the best performing models were the larger ones in the GPT-4 family, specifically GPT-4 turbo and GPT-4-o. Interestingly, modifying the Q-07 query to include the product name from the documentation stored in the RAG database significantly improved the model's performance on this query, emphasizing the importance of specificity in the query prompt. Conversely, the Mixtral 8x22B model was the poorest performer. However, Mistral Small 7B demonstrated acceptable results in most scenarios, making it a viable option for local deployments where enhancements in performance and cost efficiency are priorities, as discussed in RQ2~\ref{sec:aiops_rq2} and RQ3~\ref{sec:aiops_rq3}.

\section{Threats to Validity}
\label{sec:threats}
\phead{Internal validity.}
As previously mentioned, our experiments were conducted exclusively using the Python variants of the LangChain~\cite{langchain} and LangGraph~\cite{langgraph} libraries. We did not evaluate the behavior or functionality of their JavaScript counterparts. Future studies could explore whether the JavaScript implementations exhibit differences in performance, compatibility, or usability compared to the Python variants.

\phead{External validity.}
Threats to external validity relate to the generalizability of our findings. Our experiments were conducted using one family of frameworks for Large Language Model integration: LangChain~\cite{langchain} and LangGraph~\cite{langgraph}. These libraries abstract the implementation details necessary to interact with specific LLMs, utilizing the SDKs provided by the model providers. However, other frameworks, such as LLamaIndex~\cite{llamaindex}, offer comparable functionalities and may provide additional insights. Future studies could explore these alternative frameworks to assess their effectiveness and evaluate how they compare to LangChain and LangGraph in terms of flexibility, performance, and usability.

Additionally, abstraction frameworks like LangChain and LangGraph might impose certain limitations compared to using the native SDKs of specific models. These frameworks are designed to provide a unified interface across multiple models, which can sometimes lead to the omission of unique features available in model-specific SDKs. Consequently, failing to leverage the specialized functionalities offered by dedicated SDKs could impact the overall performance and efficiency of the agent. Future research could explore the trade-offs between using abstraction frameworks and native SDKs to optimize agent capabilities.

\phead{Construct validity.}
\label{sec:threats_construct}
Our experiments utilized specific versions of the LangChain~\cite{langchain} and LangGraph~\cite{langgraph} libraries, namely \textit{0.2.12} and \textit{0.2.10}, respectively. It is important to note that newer versions of these libraries may exhibit different behaviors, potentially incorporating improvements to the client-side handling of LLM instances. Additionally, our experimental setup depended on the serving capabilities provided by the LLM model providers (OpenAI, Anthropic, MistralAI) and their assurances regarding runtime compatibility with the integration libraries used in our study.

As Anthropic and OpenAI models are fully closed, we were unable to evaluate their agentic functionality in custom, locally deployed runtimes. While it is technically possible to test the MistralAI model family in a local environment, the high compute requirements (both CPU and GPU) for larger models rendered this option cost-prohibitive. However, we conducted tests with the Mistral Small 7B model locally using a vLLM~\cite{vllm} runtime. Unfortunately, we were unable to replicate the results obtained from the MistralAI endpoints. This inconsistency arose from a missing feature in the vLLM-LangChain integration library at the time of our experiments, specifically the absence of the \textit{bind\_tools} function. Future studies may address these limitations by exploring other runtime servers or re-evaluating the vLLM serving runtimes once the missing functionality becomes available.
\section{Related Work}
\label{sec:relatedWork}

The integration of Large Language Models (LLMs) into Artificial Intelligence for IT Operations (AIOps) represents a transformative advancement in the management and maintenance of IT systems. Models such as OpenAI's GPT-4 and Anthropic's Claude exhibit remarkable proficiency in comprehending and generating human-like text, making them valuable tools for enhancing numerous aspects of IT operations.

\noindent There are different research areas on utilizing LLMs for AIOps. One prominent area focuses on log analysis, encompassing subfields such as log parsing~\cite{Ma2024LLMParserAE, Jiang2023ALE, Xu2024DivLogLP, Ma2024OpenLogParserUP}, log anomaly detection~\cite{Liu2023InterpretableOL, Hadadi2024AnomalyDO}, and logging statement generation~\cite{Li2024GoSC, Xu2024UniLogAL}. IT systems generate vast quantities of unstructured and complex log data, and researchers are investigating how Large Language Models (LLMs) can process and interpret this data to uncover patterns, detect anomalies, and predict potential system failures. By understanding the context embedded within log messages, LLMs can proactively identify issues before they escalate, thereby minimizing downtime and enhancing system reliability. Unlike these studies, our research shifts focus from log analysis to addressing remediation workflows and procedures, highlighting a distinct contribution to the AIOps landscape.

\noindent Another research path focuses on automating incident management and response.
Studies in this area~\cite{Guo2023OWLAL, Xpert10.1145/3597503.3639081, Yu2024MonitorAssistantSC, Sarda2024LeveragingLL} demonstrate how Large Language Models (LLMs) can be leveraged to interpret alerts, correlate events, and recommend remediation steps. LLMs can also generate incident reports, summarize critical findings, and automate communication between IT teams. This capability not only expedites the incident resolution process but also reduces the cognitive load on IT staff, enabling them to dedicate more time to strategic tasks. While our work is applicable to incident management, the agents we develop—with their integrated toolsets—are designed for broader use, including preventive maintenance. Furthermore, to the best of our knowledge, our research is the first to combine predictive machine learning models with LLMs in AIOps, specifically addressing capacity planning challenges.

\section{Conclusions}
\label{sec:conclusions}

The advent of Large Language Models (LLMs) introduces transformative opportunities for managing large-scale systems. Incorporating LLM-powered agents equipped with tools significantly reduces operational burdens. When paired with predictive machine learning (ML) tools, these agents revolutionize modern IT operations and business processes. They merge the natural language understanding and contextual reasoning capabilities of LLMs with the advanced data analysis and forecasting strengths of predictive ML models. This powerful combination enables them not only to interpret complex queries but also provide proactive, actionable insights and predictions to maintain service level agreements (SLAs). By integrating predictive tools, these agents shift from reactive responses to a forward-thinking approach, improving decision making, minimizing downtime, optimizing resource utilization, and improving overall operational efficiency.

Furthermore, during anonymous surveys conducted at various events showcasing AIOps capabilities for managing Kubernetes-based infrastructure using an agentic approach with large language models, respondents expressed positive feedback. The concept was well received by diverse groups in the software industry, including development, infrastructure, and operations teams. This feedback aligns with the growing industry trend towards adopting AI-based agentic approaches to operational task management. The code and experimental results of this study are publicly available in the GitHub repository~\cite{llm-eval-aiops-code}.

Although we have proposed methods to improve the efficiency of various IT operations management processes, several challenges persist that require further investigation and exploration:
    \textbf{1) Evaluating Agentic frameworks:} We previously mentioned that our LLM-powered agents were implemented using one family of frameworks: LangChain~\cite{langchain} and LangGraph~\cite{langgraph}. While these libraries support a range of models, alternative libraries, such as LLamaIndex~\cite{llamaindex}, offer comparable functionalities and merit further exploration. Additionally, developing agents using native SDKs could present distinct advantages over generic abstraction frameworks like LangChain and LangGraph. By utilizing specialized functionalities available in native SDKs, performance and flexibility might be enhanced, providing benefits beyond the generalized features offered by abstraction frameworks.
    \textbf{2) Improving Cost Efficiency for Agentic AIOps:} As presented in Section~\ref{sec:aiops_rq3}, the use of LLM-powered agents in IT Operations management can be costly, both in terms of the time required to generate responses and the financial expenses tied to token usage. With the advancement of LLMs and the growing availability of open-source models, it is worthwhile to examine the performance of these agents when deployed locally on the same platform (e.g., RedHat OpenShift~\cite{openshift}) where tools, data, and other managed workloads reside. The local deployment of LLMs has the potential to achieve significant operational expenditure (Opex) savings. This cost reduction could facilitate the testing and implementation of autonomous monitoring agents capable of continuous 24x7 monitoring of application ecosystems. These agents could also leverage tools to perform preventive maintenance, such as real-time parameter adjustments (e.g., integrating MLOLET~\cite{MLOLET} and MLASP~\cite{MLASP} approaches) and recovery actions for monitored applications. This strategy could expand the accessibility and scalability of agents for AIOps applications, improving operational efficiency while lowering costs.

In this paper, we discuss the use of Large Language Models for adding AIOps Capabilities to Large-Scale Systems. Our contributions are:
\begin{enumerate}
\item We perform an empirical study that evaluates the effectiveness of LLM-powered agents in executing IT Operations management tasks, including capacity planning. The study examines the agents' performance based on their accuracy in resolving user queries, the time required to address these queries, and their associated costs, measured in token usage. 
\item We compare the performance of various state-of-the-art models across diverse scenarios. Our findings reveal that OpenAI's fourth-generation GPT family consistently achieves the best results within the proposed testing framework. However, other models also exhibit acceptable performance, contingent on the specific scenario and business requirements.
\item We provide recommendations for selecting appropriate LLMs and tools tailored to meet specific business requirements and challenges. Furthermore, we propose directions for future research to overcome the limitations identified in our experiments, with the goal of improving the integration and effectiveness of LLM-powered solutions in IT operations management.
\end{enumerate}

\balance

\bibliographystyle{ACM-Reference-Format}
\bibliography{aiops}


\begin{thebibliography}{39}


\ifx \showCODEN    \undefined \def \showCODEN     #1{\unskip}     \fi
\ifx \showDOI      \undefined \def \showDOI       #1{#1}\fi
\ifx \showISBNx    \undefined \def \showISBNx     #1{\unskip}     \fi
\ifx \showISBNxiii \undefined \def \showISBNxiii  #1{\unskip}     \fi
\ifx \showISSN     \undefined \def \showISSN      #1{\unskip}     \fi
\ifx \showLCCN     \undefined \def \showLCCN      #1{\unskip}     \fi
\ifx \shownote     \undefined \def \shownote      #1{#1}          \fi
\ifx \showarticletitle \undefined \def \showarticletitle #1{#1}   \fi
\ifx \showURL      \undefined \def \showURL       {\relax}        \fi
\providecommand\bibfield[2]{#2}
\providecommand\bibinfo[2]{#2}
\providecommand\natexlab[1]{#1}
\providecommand\showeprint[2][]{arXiv:#2}

\bibitem[\protect\citeauthoryear{AIOps}{AIOps}{2016}]%
        {GartnerAIOps}
AIOps \bibinfo{year}{2016}\natexlab{}.
\newblock \bibinfo{title}{Applying AIOps Platforms to Broader Datasets Will Create Unique Business Insights}.
\newblock
\newblock
\newblock
\shownote{https://www.gartner.com/en/documents/3364418.}


\bibitem[\protect\citeauthoryear{Chen, Syer, Shang, Jiang, Hassan, Nasser, and Flora}{Chen et~al\mbox{.}}{2017}]%
        {chen2017analytics}
\bibfield{author}{\bibinfo{person}{Tse-Hsun Chen}, \bibinfo{person}{Mark~D Syer}, \bibinfo{person}{Weiyi Shang}, \bibinfo{person}{Zhen~Ming Jiang}, \bibinfo{person}{Ahmed~E Hassan}, \bibinfo{person}{Mohamed Nasser}, {and} \bibinfo{person}{Parminder Flora}.} \bibinfo{year}{2017}\natexlab{}.
\newblock \showarticletitle{Analytics-driven load testing: An industrial experience report on load testing of large-scale systems}. In \bibinfo{booktitle}{\emph{2017 IEEE/ACM 39th International Conference on Software Engineering: Software Engineering in Practice Track (ICSE-SEIP)}}. \bibinfo{pages}{243--252}.
\newblock


\bibitem[\protect\citeauthoryear{Cheng, Sahoo, Saha, Yang, Liu, Woo, Singh, Saverese, and Hoi}{Cheng et~al\mbox{.}}{2023}]%
        {cheng2023ai}
\bibfield{author}{\bibinfo{person}{Qian Cheng}, \bibinfo{person}{Doyen Sahoo}, \bibinfo{person}{Amrita Saha}, \bibinfo{person}{Wenzhuo Yang}, \bibinfo{person}{Chenghao Liu}, \bibinfo{person}{Gerald Woo}, \bibinfo{person}{Manpreet Singh}, \bibinfo{person}{Silvio Saverese}, {and} \bibinfo{person}{Steven~CH Hoi}.} \bibinfo{year}{2023}\natexlab{}.
\newblock \showarticletitle{Ai for it operations (aiops) on cloud platforms: Reviews, opportunities and challenges}.
\newblock \bibinfo{journal}{\emph{arXiv preprint arXiv:2304.04661}} (\bibinfo{year}{2023}).
\newblock


\bibitem[\protect\citeauthoryear{Chu, Chen, Chen, Yu, He, Wang, Peng, Liu, Qin, and Liu}{Chu et~al\mbox{.}}{2023}]%
        {chu2023survey}
\bibfield{author}{\bibinfo{person}{Zheng Chu}, \bibinfo{person}{Jingchang Chen}, \bibinfo{person}{Qianglong Chen}, \bibinfo{person}{Weijiang Yu}, \bibinfo{person}{Tao He}, \bibinfo{person}{Haotian Wang}, \bibinfo{person}{Weihua Peng}, \bibinfo{person}{Ming Liu}, \bibinfo{person}{Bing Qin}, {and} \bibinfo{person}{Ting Liu}.} \bibinfo{year}{2023}\natexlab{}.
\newblock \showarticletitle{A survey of chain of thought reasoning: Advances, frontiers and future}.
\newblock \bibinfo{journal}{\emph{arXiv preprint arXiv:2309.15402}} (\bibinfo{year}{2023}).
\newblock


\bibitem[\protect\citeauthoryear{Edge, Trinh, Cheng, Bradley, Chao, Mody, Truitt, and Larson}{Edge et~al\mbox{.}}{2024}]%
        {Edge2024FromLT}
\bibfield{author}{\bibinfo{person}{Darren Edge}, \bibinfo{person}{Ha Trinh}, \bibinfo{person}{Newman Cheng}, \bibinfo{person}{Joshua Bradley}, \bibinfo{person}{Alex Chao}, \bibinfo{person}{Apurva Mody}, \bibinfo{person}{Steven Truitt}, {and} \bibinfo{person}{Jonathan Larson}.} \bibinfo{year}{2024}\natexlab{}.
\newblock \showarticletitle{From Local to Global: A Graph RAG Approach to Query-Focused Summarization}.
\newblock \bibinfo{journal}{\emph{ArXiv}}  \bibinfo{volume}{abs/2404.16130} (\bibinfo{year}{2024}).
\newblock
\urldef\tempurl%
\url{https://api.semanticscholar.org/CorpusID:269363075}
\showURL{%
\tempurl}


\bibitem[\protect\citeauthoryear{Evaluating LLM Agents for AIOps on Red Hat OpenShift}{Evaluating LLM Agents for AIOps on Red Hat OpenShift}{2024}]%
        {llm-eval-aiops-code}
Evaluating LLM Agents for AIOps on Red Hat OpenShift \bibinfo{year}{2024}\natexlab{}.
\newblock \bibinfo{title}{Evaluating LLM Agents for AIOps on Red Hat OpenShift - A full implementation example of agents with LangGraph and LangChain on Red Hat OpenShift}.
\newblock
\newblock
\newblock
\shownote{https://github.com/eartvit/llm-agents-on-ocp.}


\bibitem[\protect\citeauthoryear{Goswami}{Goswami}{[n.d.]}]%
        {goswamichallenges}
\bibfield{author}{\bibinfo{person}{Maloy~Jyoti Goswami}.} \bibinfo{year}{[n.d.]}\natexlab{}.
\newblock \showarticletitle{Challenges and Solutions in Integrating AI with Multi-Cloud Architectures}.
\newblock  (\bibinfo{year}{[n.\,d.]}).
\newblock


\bibitem[\protect\citeauthoryear{Guo, Yang, Liu, Yang, Chai, Bai, Peng, Hu, Chen, Zhang, Shi, Zheng, Zheng, Zhang, Xu, and Li}{Guo et~al\mbox{.}}{2023}]%
        {Guo2023OWLAL}
\bibfield{author}{\bibinfo{person}{Hongcheng Guo}, \bibinfo{person}{Jian Yang}, \bibinfo{person}{Jiaheng Liu}, \bibinfo{person}{Liqun Yang}, \bibinfo{person}{Linzheng Chai}, \bibinfo{person}{Jiaqi Bai}, \bibinfo{person}{Junran Peng}, \bibinfo{person}{Xiaorong Hu}, \bibinfo{person}{Chao Chen}, \bibinfo{person}{Dongfeng Zhang}, \bibinfo{person}{Xu Shi}, \bibinfo{person}{Tieqiao Zheng}, \bibinfo{person}{Liangfan Zheng}, \bibinfo{person}{Bo Zhang}, \bibinfo{person}{Ke Xu}, {and} \bibinfo{person}{Zhoujun Li}.} \bibinfo{year}{2023}\natexlab{}.
\newblock \showarticletitle{OWL: A Large Language Model for IT Operations}.
\newblock \bibinfo{journal}{\emph{ArXiv}}  \bibinfo{volume}{abs/2309.09298} (\bibinfo{year}{2023}).
\newblock
\urldef\tempurl%
\url{https://api.semanticscholar.org/CorpusID:262043747}
\showURL{%
\tempurl}


\bibitem[\protect\citeauthoryear{Hadadi, Xu, Bianculli, and Briand}{Hadadi et~al\mbox{.}}{2024}]%
        {Hadadi2024AnomalyDO}
\bibfield{author}{\bibinfo{person}{Fateme Hadadi}, \bibinfo{person}{Qinghua Xu}, \bibinfo{person}{Domenico Bianculli}, {and} \bibinfo{person}{Lionel~C. Briand}.} \bibinfo{year}{2024}\natexlab{}.
\newblock \showarticletitle{Anomaly Detection on Unstable Logs with GPT Models}.
\newblock \bibinfo{journal}{\emph{ArXiv}}  \bibinfo{volume}{abs/2406.07467} (\bibinfo{year}{2024}).
\newblock
\urldef\tempurl%
\url{https://api.semanticscholar.org/CorpusID:270379599}
\showURL{%
\tempurl}


\bibitem[\protect\citeauthoryear{Hedström, Weber, Bareeva, Krakowczyk, Motzkus, Samek, Lapuschkin, and Höhne}{Hedström et~al\mbox{.}}{2023}]%
        {hedström2023quantusexplainableaitoolkit}
\bibfield{author}{\bibinfo{person}{Anna Hedström}, \bibinfo{person}{Leander Weber}, \bibinfo{person}{Dilyara Bareeva}, \bibinfo{person}{Daniel Krakowczyk}, \bibinfo{person}{Franz Motzkus}, \bibinfo{person}{Wojciech Samek}, \bibinfo{person}{Sebastian Lapuschkin}, {and} \bibinfo{person}{Marina M.~C. Höhne}.} \bibinfo{year}{2023}\natexlab{}.
\newblock \bibinfo{title}{Quantus: An Explainable AI Toolkit for Responsible Evaluation of Neural Network Explanations and Beyond}.
\newblock
\newblock
\showeprint[arxiv]{2202.06861}~[cs.LG]
\urldef\tempurl%
\url{https://arxiv.org/abs/2202.06861}
\showURL{%
\tempurl}


\bibitem[\protect\citeauthoryear{Jiang, Zhang, He, Yang, Ma, Qin, Kang, Dang, Rajmohan, Lin, and Zhang}{Jiang et~al\mbox{.}}{2024}]%
        {Xpert10.1145/3597503.3639081}
\bibfield{author}{\bibinfo{person}{Yuxuan Jiang}, \bibinfo{person}{Chaoyun Zhang}, \bibinfo{person}{Shilin He}, \bibinfo{person}{Zhihao Yang}, \bibinfo{person}{Minghua Ma}, \bibinfo{person}{Si Qin}, \bibinfo{person}{Yu Kang}, \bibinfo{person}{Yingnong Dang}, \bibinfo{person}{Saravan Rajmohan}, \bibinfo{person}{Qingwei Lin}, {and} \bibinfo{person}{Dongmei Zhang}.} \bibinfo{year}{2024}\natexlab{}.
\newblock \showarticletitle{Xpert: Empowering Incident Management with Query Recommendations via Large Language Models}. In \bibinfo{booktitle}{\emph{Proceedings of the IEEE/ACM 46th International Conference on Software Engineering}} (Lisbon, Portugal) \emph{(\bibinfo{series}{ICSE '24})}. \bibinfo{publisher}{Association for Computing Machinery}, \bibinfo{address}{New York, NY, USA}, Article \bibinfo{articleno}{92}, \bibinfo{numpages}{13}~pages.
\newblock
\showISBNx{9798400702174}
\urldef\tempurl%
\url{https://doi.org/10.1145/3597503.3639081}
\showDOI{\tempurl}


\bibitem[\protect\citeauthoryear{Jiang, Liu, Huang, Li, Huo, Gu, Chen, Zhu, and Lyu}{Jiang et~al\mbox{.}}{2023}]%
        {Jiang2023ALE}
\bibfield{author}{\bibinfo{person}{Zhihan Jiang}, \bibinfo{person}{Jinyang Liu}, \bibinfo{person}{Junjie Huang}, \bibinfo{person}{Yichen Li}, \bibinfo{person}{Yintong Huo}, \bibinfo{person}{Jia-Yuan Gu}, \bibinfo{person}{Zhuangbin Chen}, \bibinfo{person}{Jieming Zhu}, {and} \bibinfo{person}{Michael~R. Lyu}.} \bibinfo{year}{2023}\natexlab{}.
\newblock \showarticletitle{A Large-Scale Evaluation for Log Parsing Techniques: How Far Are We?}
\newblock \bibinfo{journal}{\emph{Proceedings of the 33rd ACM SIGSOFT International Symposium on Software Testing and Analysis}} (\bibinfo{year}{2023}).
\newblock
\urldef\tempurl%
\url{https://api.semanticscholar.org/CorpusID:261049500}
\showURL{%
\tempurl}


\bibitem[\protect\citeauthoryear{Kohlbrenner, Bauer, Nakajima, Binder, Samek, and Lapuschkin}{Kohlbrenner et~al\mbox{.}}{2020}]%
        {kohlbrenner2020bestpracticeexplainingneural}
\bibfield{author}{\bibinfo{person}{Maximilian Kohlbrenner}, \bibinfo{person}{Alexander Bauer}, \bibinfo{person}{Shinichi Nakajima}, \bibinfo{person}{Alexander Binder}, \bibinfo{person}{Wojciech Samek}, {and} \bibinfo{person}{Sebastian Lapuschkin}.} \bibinfo{year}{2020}\natexlab{}.
\newblock \bibinfo{title}{Towards Best Practice in Explaining Neural Network Decisions with LRP}.
\newblock
\newblock
\showeprint[arxiv]{1910.09840}~[cs.LG]
\urldef\tempurl%
\url{https://arxiv.org/abs/1910.09840}
\showURL{%
\tempurl}


\bibitem[\protect\citeauthoryear{Kubernetes}{Kubernetes}{2022}]%
        {kubernetes}
Kubernetes \bibinfo{year}{2022}\natexlab{}.
\newblock \bibinfo{title}{Kubernetes - Production-Grade Container Orchestration}.
\newblock
\newblock
\newblock
\shownote{https://kubernetes.io/.}


\bibitem[\protect\citeauthoryear{Kumar}{Kumar}{2022}]%
        {kumar2022challenges}
\bibfield{author}{\bibinfo{person}{Bharath Kumar}.} \bibinfo{year}{2022}\natexlab{}.
\newblock \showarticletitle{Challenges and Solutions for Integrating AI with Multi-Cloud Architectures}.
\newblock \bibinfo{journal}{\emph{International Journal of Multidisciplinary Innovation and Research Methodology, ISSN: 2960-2068}} \bibinfo{volume}{1}, \bibinfo{number}{1} (\bibinfo{year}{2022}), \bibinfo{pages}{71--77}.
\newblock


\bibitem[\protect\citeauthoryear{LangChain}{LangChain}{2024}]%
        {langchain}
LangChain \bibinfo{year}{2024}\natexlab{}.
\newblock \bibinfo{title}{LangChain - a framework for developing applications powered by large language models (LLMs).}
\newblock
\newblock
\newblock
\shownote{https://python.langchain.com/docs/introduction/.}


\bibitem[\protect\citeauthoryear{LangGraph}{LangGraph}{2024}]%
        {langgraph}
LangGraph \bibinfo{year}{2024}\natexlab{}.
\newblock \bibinfo{title}{LangGraph - Building Agents as Graphs}.
\newblock
\newblock
\newblock
\shownote{https://langchain-ai.github.io/langgraph/.}


\bibitem[\protect\citeauthoryear{Levin, Garion, Kolodner, Lorenz, Barabash, Kugler, and McShane}{Levin et~al\mbox{.}}{2019}]%
        {Levin_2019}
\bibfield{author}{\bibinfo{person}{Anna Levin}, \bibinfo{person}{Shelly Garion}, \bibinfo{person}{Elliot~K. Kolodner}, \bibinfo{person}{Dean~H. Lorenz}, \bibinfo{person}{Katherine Barabash}, \bibinfo{person}{Mike Kugler}, {and} \bibinfo{person}{Niall McShane}.} \bibinfo{year}{2019}\natexlab{}.
\newblock \showarticletitle{AIOps for a Cloud Object Storage Service}. In \bibinfo{booktitle}{\emph{2019 IEEE International Congress on Big Data (BigDataCongress)}}. \bibinfo{publisher}{IEEE}, \bibinfo{pages}{165–169}.
\newblock
\urldef\tempurl%
\url{https://doi.org/10.1109/bigdatacongress.2019.00036}
\showDOI{\tempurl}


\bibitem[\protect\citeauthoryear{Li, Huo, Zhong, Jiang, Liu, Huang, Gu, He, and Lyu}{Li et~al\mbox{.}}{2024}]%
        {Li2024GoSC}
\bibfield{author}{\bibinfo{person}{Yichen Li}, \bibinfo{person}{Yintong Huo}, \bibinfo{person}{Renyi Zhong}, \bibinfo{person}{Zhihan Jiang}, \bibinfo{person}{Jinyang Liu}, \bibinfo{person}{Junjie Huang}, \bibinfo{person}{Jiazhen Gu}, \bibinfo{person}{Pinjia He}, {and} \bibinfo{person}{Michael~R. Lyu}.} \bibinfo{year}{2024}\natexlab{}.
\newblock \showarticletitle{Go Static: Contextualized Logging Statement Generation}.
\newblock \bibinfo{journal}{\emph{ArXiv}}  \bibinfo{volume}{abs/2402.12958} (\bibinfo{year}{2024}).
\newblock
\urldef\tempurl%
\url{https://api.semanticscholar.org/CorpusID:267760100}
\showURL{%
\tempurl}


\bibitem[\protect\citeauthoryear{Liu, Tao, Meng, Wang, Ma, Chen, Zhao, Yang, and Jiang}{Liu et~al\mbox{.}}{2023}]%
        {Liu2023InterpretableOL}
\bibfield{author}{\bibinfo{person}{Yilun Liu}, \bibinfo{person}{Shimin Tao}, \bibinfo{person}{Weibin Meng}, \bibinfo{person}{Jingyu Wang}, \bibinfo{person}{Wenbing Ma}, \bibinfo{person}{Yuhang Chen}, \bibinfo{person}{Yanqing Zhao}, \bibinfo{person}{Hao Yang}, {and} \bibinfo{person}{Yanfei Jiang}.} \bibinfo{year}{2023}\natexlab{}.
\newblock \showarticletitle{Interpretable Online Log Analysis Using Large Language Models with Prompt Strategies}.
\newblock \bibinfo{journal}{\emph{2024 IEEE/ACM 32nd International Conference on Program Comprehension (ICPC)}} (\bibinfo{year}{2023}), \bibinfo{pages}{35--46}.
\newblock
\urldef\tempurl%
\url{https://api.semanticscholar.org/CorpusID:260900274}
\showURL{%
\tempurl}


\bibitem[\protect\citeauthoryear{LlamaIndex}{LlamaIndex}{2024}]%
        {llamaindex}
LlamaIndex \bibinfo{year}{2024}\natexlab{}.
\newblock \bibinfo{title}{LlamaIndex - a framework for building context-augmented generative AI applications with LLMs including agents and workflows.}
\newblock
\newblock
\newblock
\shownote{https://docs.llamaindex.ai/en/stable/.}


\bibitem[\protect\citeauthoryear{Lwakatare, Raj, Crnkovic, Bosch, and Olsson}{Lwakatare et~al\mbox{.}}{2020}]%
        {LWAKATARE2020106368}
\bibfield{author}{\bibinfo{person}{Lucy~Ellen Lwakatare}, \bibinfo{person}{Aiswarya Raj}, \bibinfo{person}{Ivica Crnkovic}, \bibinfo{person}{Jan Bosch}, {and} \bibinfo{person}{Helena~Holmström Olsson}.} \bibinfo{year}{2020}\natexlab{}.
\newblock \showarticletitle{Large-scale machine learning systems in real-world industrial settings: A review of challenges and solutions}.
\newblock \bibinfo{journal}{\emph{Information and Software Technology}}  \bibinfo{volume}{127} (\bibinfo{year}{2020}), \bibinfo{pages}{106368}.
\newblock
\showISSN{0950-5849}
\urldef\tempurl%
\url{https://doi.org/10.1016/j.infsof.2020.106368}
\showDOI{\tempurl}


\bibitem[\protect\citeauthoryear{Ma, Chen, Kim, Chen, and Wang}{Ma et~al\mbox{.}}{2024a}]%
        {Ma2024LLMParserAE}
\bibfield{author}{\bibinfo{person}{Zeyang Ma}, \bibinfo{person}{An~Ran Chen}, \bibinfo{person}{Dong~Jae Kim}, \bibinfo{person}{Tse-Husn Chen}, {and} \bibinfo{person}{Shaowei Wang}.} \bibinfo{year}{2024}\natexlab{a}.
\newblock \showarticletitle{LLMParser: An Exploratory Study on Using Large Language Models for Log Parsing}.
\newblock \bibinfo{journal}{\emph{2024 IEEE/ACM 46th International Conference on Software Engineering (ICSE)}} (\bibinfo{year}{2024}), \bibinfo{pages}{1209--1221}.
\newblock
\urldef\tempurl%
\url{https://api.semanticscholar.org/CorpusID:269123285}
\showURL{%
\tempurl}


\bibitem[\protect\citeauthoryear{Ma, Kim, and Chen}{Ma et~al\mbox{.}}{2024b}]%
        {Ma2024OpenLogParserUP}
\bibfield{author}{\bibinfo{person}{Zeyang Ma}, \bibinfo{person}{Dong~Jae Kim}, {and} \bibinfo{person}{Tse-Husn Chen}.} \bibinfo{year}{2024}\natexlab{b}.
\newblock \showarticletitle{OpenLogParser: Unsupervised Parsing with Open-Source Large Language Models}.
\newblock \bibinfo{journal}{\emph{ArXiv}}  \bibinfo{volume}{abs/2408.01585} (\bibinfo{year}{2024}).
\newblock
\urldef\tempurl%
\url{https://api.semanticscholar.org/CorpusID:271709638}
\showURL{%
\tempurl}


\bibitem[\protect\citeauthoryear{Prometheus}{Prometheus}{2022}]%
        {prometheus}
Prometheus \bibinfo{year}{2022}\natexlab{}.
\newblock \bibinfo{title}{Prometheus - Monitoring System and Timeseries Database}.
\newblock
\newblock
\newblock
\shownote{https://prometheus.io/.}


\bibitem[\protect\citeauthoryear{Pydantic}{Pydantic}{2024}]%
        {pydantic}
Pydantic \bibinfo{year}{2024}\natexlab{}.
\newblock \bibinfo{title}{Pydantic - a widely used data validation library for Python.}
\newblock
\newblock
\newblock
\shownote{https://docs.pydantic.dev/latest/.}


\bibitem[\protect\citeauthoryear{Red Hat OpenShift}{Red Hat OpenShift}{2022}]%
        {openshift}
Red Hat OpenShift \bibinfo{year}{2022}\natexlab{}.
\newblock \bibinfo{title}{Red Hat OpenShift - the industry's leading hybrid cloud application platform powered by Kubernetes}.
\newblock
\newblock
\newblock
\shownote{https://www.redhat.com/en/technologies/cloud-computing/openshift.}


\bibitem[\protect\citeauthoryear{Sai, Mohankumar, and Khapra}{Sai et~al\mbox{.}}{2020}]%
        {sai2020surveyevaluationmetricsused}
\bibfield{author}{\bibinfo{person}{Ananya~B. Sai}, \bibinfo{person}{Akash~Kumar Mohankumar}, {and} \bibinfo{person}{Mitesh~M. Khapra}.} \bibinfo{year}{2020}\natexlab{}.
\newblock \bibinfo{title}{A Survey of Evaluation Metrics Used for NLG Systems}.
\newblock
\newblock
\showeprint[arxiv]{2008.12009}~[cs.CL]
\urldef\tempurl%
\url{https://arxiv.org/abs/2008.12009}
\showURL{%
\tempurl}


\bibitem[\protect\citeauthoryear{Sarda, Namrud, Litoiu, Shwartz, and Watts}{Sarda et~al\mbox{.}}{2024}]%
        {Sarda2024LeveragingLL}
\bibfield{author}{\bibinfo{person}{Komal Sarda}, \bibinfo{person}{Zakeya Namrud}, \bibinfo{person}{Marin Litoiu}, \bibinfo{person}{Larisa Shwartz}, {and} \bibinfo{person}{Ian Watts}.} \bibinfo{year}{2024}\natexlab{}.
\newblock \showarticletitle{Leveraging Large Language Models for the Auto-remediation of Microservice Applications: An Experimental Study}. In \bibinfo{booktitle}{\emph{SIGSOFT FSE Companion}}.
\newblock
\urldef\tempurl%
\url{https://api.semanticscholar.org/CorpusID:271099029}
\showURL{%
\tempurl}


\bibitem[\protect\citeauthoryear{Vitui and Chen}{Vitui and Chen}{2021}]%
        {MLASP}
\bibfield{author}{\bibinfo{person}{Arthur Vitui} {and} \bibinfo{person}{Tse-Hsun~Peter Chen}.} \bibinfo{year}{2021}\natexlab{}.
\newblock \showarticletitle{MLASP: Machine learning assisted capacity planning}.
\newblock \bibinfo{journal}{\emph{Empirical Software Engineering}}  \bibinfo{volume}{26} (\bibinfo{year}{2021}), \bibinfo{pages}{1--27}.
\newblock
\urldef\tempurl%
\url{https://api.semanticscholar.org/CorpusID:236930377}
\showURL{%
\tempurl}


\bibitem[\protect\citeauthoryear{Vitui and Chen}{Vitui and Chen}{2024}]%
        {MLOLET}
\bibfield{author}{\bibinfo{person}{Arthur Vitui} {and} \bibinfo{person}{Tse-Hsun~Peter Chen}.} \bibinfo{year}{2024}\natexlab{}.
\newblock \showarticletitle{MLOLET - Machine Learning Optimized Load and Endurance Testing: An industrial experience report}. In \bibinfo{booktitle}{\emph{Proceedings of the 39th IEEE/ACM International Conference on Automated Software Engineering}} (Sacramento, CA, USA) \emph{(\bibinfo{series}{ASE’24})}. \bibinfo{publisher}{IEEE Press}, \bibinfo{pages}{465474}.
\newblock
\showISBNx{979-8-4007-1248-7/24/10}


\bibitem[\protect\citeauthoryear{vLLM}{vLLM}{2024}]%
        {vllm}
vLLM \bibinfo{year}{2024}\natexlab{}.
\newblock \bibinfo{title}{vLLM - a fast and easy-to-use library for LLM inference and serving.}
\newblock
\newblock
\newblock
\shownote{https://docs.vllm.ai/en/latest/.}


\bibitem[\protect\citeauthoryear{Wei, Wang, Schuurmans, Bosma, Ichter, Xia, Chi, Le, and Zhou}{Wei et~al\mbox{.}}{2023}]%
        {wei2023chainofthoughtpromptingelicitsreasoning}
\bibfield{author}{\bibinfo{person}{Jason Wei}, \bibinfo{person}{Xuezhi Wang}, \bibinfo{person}{Dale Schuurmans}, \bibinfo{person}{Maarten Bosma}, \bibinfo{person}{Brian Ichter}, \bibinfo{person}{Fei Xia}, \bibinfo{person}{Ed Chi}, \bibinfo{person}{Quoc Le}, {and} \bibinfo{person}{Denny Zhou}.} \bibinfo{year}{2023}\natexlab{}.
\newblock \bibinfo{title}{Chain-of-Thought Prompting Elicits Reasoning in Large Language Models}.
\newblock
\newblock
\showeprint[arxiv]{2201.11903}~[cs.CL]
\urldef\tempurl%
\url{https://arxiv.org/abs/2201.11903}
\showURL{%
\tempurl}


\bibitem[\protect\citeauthoryear{WireMock}{WireMock}{2022}]%
        {wiremock}
WireMock \bibinfo{year}{2022}\natexlab{}.
\newblock \bibinfo{title}{WireMock: Mock the APIs You Depend On}.
\newblock
\newblock
\newblock
\shownote{https://wiremock.org/.}


\bibitem[\protect\citeauthoryear{WireMock Metrics}{WireMock Metrics}{2023}]%
        {wiremock-metrics2}
WireMock Metrics \bibinfo{year}{2023}\natexlab{}.
\newblock \bibinfo{title}{WireMock Metrics - extended WireMock with Prometheus metrics and global random string payload ResponseTransformer}.
\newblock
\newblock
\newblock
\shownote{https://github.com/eartvit/wiremock-metrics2.}


\bibitem[\protect\citeauthoryear{Xu, Cui, Zhao, Zhang, He, He, Li, Kang, Lin, Dang, Rajmohan, and Zhang}{Xu et~al\mbox{.}}{2024a}]%
        {Xu2024UniLogAL}
\bibfield{author}{\bibinfo{person}{Junjielong Xu}, \bibinfo{person}{Ziang Cui}, \bibinfo{person}{Yuan Zhao}, \bibinfo{person}{Xu Zhang}, \bibinfo{person}{Shilin He}, \bibinfo{person}{Pinjia He}, \bibinfo{person}{Liqun Li}, \bibinfo{person}{Yu Kang}, \bibinfo{person}{Qingwei Lin}, \bibinfo{person}{Yingnong Dang}, \bibinfo{person}{S. Rajmohan}, {and} \bibinfo{person}{Dongmei Zhang}.} \bibinfo{year}{2024}\natexlab{a}.
\newblock \showarticletitle{UniLog: Automatic Logging via LLM and In-Context Learning}.
\newblock \bibinfo{journal}{\emph{2024 IEEE/ACM 46th International Conference on Software Engineering (ICSE)}} (\bibinfo{year}{2024}), \bibinfo{pages}{1--12}.
\newblock
\urldef\tempurl%
\url{https://api.semanticscholar.org/CorpusID:267523731}
\showURL{%
\tempurl}


\bibitem[\protect\citeauthoryear{Xu, Yang, Huo, Zhang, and He}{Xu et~al\mbox{.}}{2024b}]%
        {Xu2024DivLogLP}
\bibfield{author}{\bibinfo{person}{Junjielong Xu}, \bibinfo{person}{Ruichun Yang}, \bibinfo{person}{Yintong Huo}, \bibinfo{person}{Chengyu Zhang}, {and} \bibinfo{person}{Pinjia He}.} \bibinfo{year}{2024}\natexlab{b}.
\newblock \showarticletitle{DivLog: Log Parsing with Prompt Enhanced In-Context Learning}.
\newblock \bibinfo{journal}{\emph{2024 IEEE/ACM 46th International Conference on Software Engineering (ICSE)}} (\bibinfo{year}{2024}), \bibinfo{pages}{2457--2468}.
\newblock
\urldef\tempurl%
\url{https://api.semanticscholar.org/CorpusID:269123195}
\showURL{%
\tempurl}


\bibitem[\protect\citeauthoryear{Yao, Zhao, Yu, Du, Shafran, Narasimhan, and Cao}{Yao et~al\mbox{.}}{2023}]%
        {yao2023reactsynergizingreasoningacting}
\bibfield{author}{\bibinfo{person}{Shunyu Yao}, \bibinfo{person}{Jeffrey Zhao}, \bibinfo{person}{Dian Yu}, \bibinfo{person}{Nan Du}, \bibinfo{person}{Izhak Shafran}, \bibinfo{person}{Karthik Narasimhan}, {and} \bibinfo{person}{Yuan Cao}.} \bibinfo{year}{2023}\natexlab{}.
\newblock \bibinfo{title}{ReAct: Synergizing Reasoning and Acting in Language Models}.
\newblock
\newblock
\showeprint[arxiv]{2210.03629}~[cs.CL]
\urldef\tempurl%
\url{https://arxiv.org/abs/2210.03629}
\showURL{%
\tempurl}


\bibitem[\protect\citeauthoryear{Yu, Ma, Zhang, Qin, Kang, Bansal, Rajmohan, Dang, Pei, Pei, Lin, and Zhang}{Yu et~al\mbox{.}}{2024}]%
        {Yu2024MonitorAssistantSC}
\bibfield{author}{\bibinfo{person}{Zhaoyang Yu}, \bibinfo{person}{Ming-Jie Ma}, \bibinfo{person}{Chaoyun Zhang}, \bibinfo{person}{Si Qin}, \bibinfo{person}{Yu Kang}, \bibinfo{person}{Chetan Bansal}, \bibinfo{person}{S. Rajmohan}, \bibinfo{person}{Yingnong Dang}, \bibinfo{person}{Changhua Pei}, \bibinfo{person}{Dan Pei}, \bibinfo{person}{Qingwei Lin}, {and} \bibinfo{person}{Dongmei Zhang}.} \bibinfo{year}{2024}\natexlab{}.
\newblock \showarticletitle{MonitorAssistant: Simplifying Cloud Service Monitoring via Large Language Models}. In \bibinfo{booktitle}{\emph{SIGSOFT FSE Companion}}.
\newblock
\urldef\tempurl%
\url{https://api.semanticscholar.org/CorpusID:271098855}
\showURL{%
\tempurl}


\end{thebibliography}
\end{document}